\documentclass[runningheads]{llncs}
\usepackage{nicefrac,amsmath,graphicx,xcolor,amsfonts,enumerate}
\usepackage{hyperref}
\newcommand\OZ[1]{1\;\PC{\PF#1}\;0}
\newcommand\CalX {{\ensuremath{\cal X}}}
\newcommand\Body {\Meta{Body}}
\newcommand\Spec {\Meta{Spec}}
\newcommand\Imp {\Meta{Imp}}
\newcommand\Left {\Meta{Left}}
\newcommand\Right {\Meta{Right}}
\newcommand\Prog {\Meta{Prog}}
\newcommand\Prop {\Meta{prop}}
\newcommand\Pre {\Meta{pre}}
\newcommand\Post {\Meta{post}}
\newcommand\pGCL {\textit{pGCL}}
\newcommand\Exp {\Meta{E}}
\newcommand\Cond {\Meta{C}}
\newcommand\Sbst[3]{#1[#2\ensuremath\backslash#3]}
\newcommand\B {\kern.12em\raisebox{.5ex}{\framebox[.35em]{\rule{0pt}{.5ex}}}\kern.47em}
\newcommand\Ass[1] {\ensuremath{\color{blue}\left\{\,{\PF{#1}}\,\right\}}}
\newcommand\Qdot {\HangRight{\textrm{\quad.}}}
\newcommand\Qcomma {\HangRight{\textrm{\quad,}}}
\newcommand\AST {\textit{AST}}
\newcommand\In {:\kern-.1em$\in$\kern-.2em}
\newcommand\St {:\kern-.1em$|$\kern-.2em}
\newcommand\CbC {\textit{C\kern.03emb\kern-.045emC}}

\newcommand\Cite[1] {{\color{red}\sf[#1]}}

\newcommand\Itm[1] {(\ref{#1})}
\newcommand\Sec[1] {Sec.~\ref{#1}}
\newcommand\Union		{\mathbin{\cup}}
\newcommand\Ref			{\mathrel\sqsubseteq}
\newcommand\Prg[1] {Program\,(\ref{#1})}
\newcommand\LHS			{\textit{lhs}}
\newcommand\RHS			{\textit{rhs}}
\newcommand\Supp[1]	{\ensuremath{\lceil#1\rceil}} % Support of a distribution.
\newcommand\Dist {\ensuremath{\mathbb D}}
\newcommand\Pow			{{\mathbb P}}%\kern-.05em}
\newcommand\FF {\NF{1}{2}} % Shortcut: "fifty-fifty".			
\newcommand\Min {\ensuremath{\mathbin{\sf min}}}
\newcommand\WP {\textsf{wp}}
\newcommand\WPP[2] {\WP(\texttt{#1},#2)}
\newcommand\False {\ensuremath{\mathit{False}}}
\newcommand\True {\ensuremath{\mathit{True}}}
\newcommand\Meta[1] {{\texttt{\textit{#1}}}}
\newcommand\HangRight[1]	{\makebox[0pt][l]{#1}}
\newcommand\HangLeft[1]		{\makebox[0pt][r]{#1}}
\newcommand\Implies		{\mathbin{\Rightarrow}} % Proper spacing for binary operator.
\newcommand\ND			{\ensuremath{\mathbin{\sqcap}}}
\newcommand\NF[2]		{\ensuremath{\nicefrac{#1}{#2}}} % Abbreviation.
\newcommand\PC[1]		{\ensuremath{\mathbin{{}_{{#1}}\kern-.05em\oplus}}}
\newcommand\PPC[2]		{\ensuremath{\mathbin{{}_{{#1}}\kern-.05em\oplus_{{#2}}}}}
\newcommand\PCF			{\PCNF{1}{2}}
\newcommand\PCNF[2]		{\PC{\NF{#1}{#2}}}
\newcommand\PF[1][\empty]	{\ifx\empty#1\empty\tt\else\texttt{#1}\fi} % Program font: \PF sets it; \PF[x] applies it.
\newcommand\Wide[1] 		{~~~#1~~~} % Basic space.
\newcommand\WIDERM[1]	{\Wide{\WideRm{#1}}} % Extra wide space around text.
\newcommand\WideRm[1]	{~~~\textrm{#1}~~~} % Basic space, but around text.

\newcommand\Eqn[1] {(\ref{#1})}
\newcommand\Fig[2][\empty] {\ifx#1\empty Fig.~\ref{#2}\else Fig.~\ref{#2}(#1)\fi}
\newcommand\ProgIL[1] {~\texttt{#1}~}

\newenvironment{ProgEqn}[1][]
	{\def\L{#1}\ifx\L\empty\[\else\begin{equation}\label{#1}\fi%
		\begin{minipage}{.8\linewidth}\texttt\bgroup%
			\vspace*{-2.5ex} % tabbing introduces unwanted vertical space.
			\begin{tabbing}
				\quad\=\quad\=\quad\=\quad\=\quad\=\quad\=
				\quad\=\quad\=\quad\=\quad\=\quad\=\quad\=
				\quad\=
			\kill}
	{\end{tabbing}\egroup\end{minipage}\ifx\L\empty\]\else\end{equation}\fi}
						
\newcommand\Com {\`\footnotesize\sf--~} % Comment for ProgMini, right-adjusted.
\newcommand\LCom {\footnotesize\sf--~} % Comment for ProgMini, at left.

\newenvironment{Reason}{\begin{tabbing}\hspace{4em}\= \hspace{1cm} \= \kill}
{\end{tabbing}\vspace{-1em}}
\newcommand\Step[2] {#1 \> $\begin{array}[t]{@{}llll}#2\end{array}$ \\}
\newcommand\StepR[3] {#1 \> $\begin{array}[t]{@{}llll}#3\end{array}$
\` {\RF \makebox[0pt][r]{\begin{tabular}[t]{r}``#2''\end{tabular}}} \\}

\newcommand\Space {~ \\}
\newcommand\RF {\small}

\newenvironment{Refines}{\begin{list}{$\Ref$}{
\setlength{\rightmargin}{0em}
\addtolength{\leftmargin}{1em}
\setlength{\listparindent}{0em}
\setlength{\labelwidth}{2em}}}{
\end{list}}
\newcommand\Unspace {\vspace{-1.4ex}}

\begin{document}
\title{Correctness by construction \\ for probabilistic programs\thanks{We are grateful for the support of the Australian Research Council.}}
\titlerunning{Correctness by construction \\ for probabilistic programs}
\author{Annabelle McIver\inst{1}\and Carroll Morgan\inst{2}}
\institute{University of New South Wales \& Trustworthy Systems, Data61, CSIRO \\\email{carroll.morgan@unsw.edu.au}
\and Macquarie University \email{anabelle.mciver@mq.edu.au}}

\maketitle

\begin{abstract}
The ``correct by construction'' paradigm is an important component of modern Formal Methods, and here we use the probabilistic Guarded-Command Language \pGCL\ to illustrate its application to \emph{probabilistic} programming.

\quad\pGCL\ extends Dij\-kstra's guarded-command language \textit{GCL} with probabilistic choice, and is equipped with a correctness-preserving refinement relation $(\Ref)$ that enables compact, abstract specifications of probabilistic properties to
be transformed gradually to concrete, executable code by applying mathematical insights in a systematic and layered way.

\quad Characteristically for ``correctness by construction'', as far as possible the reasoning in each refinement-step layer does not depend on earlier layers, and does not affect later ones.

\quad We demonstrate the technique by deriving a fair-coin implementation of any given discrete 
probability distribution. In the special case of simulating a fair die, 
our correct-by-construction algorithm turns out to be ``within spitting distance'' of Knuth and Yao's optimal solution.
\end{abstract}

\section{Testing probabilistic programs?}

Edsger Dij\-kstra argued \cite[p3]{Dijkstra:aa} that the construction of  \emph{correct} programs requires mathematical proof, since ``\ldots program testing can be used very effectively to show the presence of bugs but never to show their absence.''
But for programs that are constructed to exhibit some form of randomisation, regular testing can't even establish that \emph{presence}:   odd program traces are almost always bound to turn up even in \emph{correctly} operating probabilistic systems.

Thus evidence of quantitative errors in probabilistic systems would require many, many traces to be subjected to detailed statistical analysis --- yet even then debugging probabilistic programs remains a challenge when that evidence has been assembled. Unlike standard (non-probabilistic programs), where a failed test can often pinpoint the source of the offending error in the code, it's not easy to figure out what to change in the implementation of probabilistic programs in order to move closer towards ``correctness'' rather than further away.  

Without that unambiguous relationship between  failed tests and the coding errors that cause them, Dij\-kstra's caution regarding proofs of programs is even more apposite. In this paper we describe such a proof method for probability:
correctness-by-construction. In a sentence, to apply ``\CbC'' one constructs the program and its proof at the same time, letting the requirement that there \emph{be} a proof guide the design decisions taken while constructing the program.

Like standard programs, probabilistic programs incorporate mathematical insights into algorithms, and a correctness-by-construction method should allow a program developer to refer rigorously to those insights by applying development steps that preserve ``probabilistic correctness''. 
Probabilistic correctness is however notoriously unintuitive. For example, the solution of the infamous Monty Hall problem caused such a ruckus in the mathematical community that even Paul Erd\"os questioned the correct analysis \cite{Vazsonyi:2002aa}.\,% 
\footnote{A game show host, Monty Hall, shows a contestant three curtains, behind one of which sits a Cadillac; the other two curtains conceal goats. The contestant guesses which curtain hides the prize, and Monty then opens another that concealed a goat. The contestant is allowed to change his mind. Should he?}
Yet once coded up as a program \cite[p22]{McIver:05a}, the Monty Hall problem is only four lines long! More generally though,  many widely relied-upon programs in security are quite short, and yet still pose significant challenges for correctness.
%\Footnote{Is this true? Check Gilles' work.}

We describe correctness-by-construction in the context of \pGCL, a small programming language which restores demonic choice to  Kozen's landmark (purely) probabilistic semantics \cite{Kozen:81,Kozen:83} while using the syntax of Dijkstra's \textit{GCL} \cite{Dijkstra:76}. Its basic principles are that correctness for programs can be described by a generalisation of Hoare logic that includes \emph{quantitative} analysis; and it has a definition of refinement that allows programs to be developed in such a way that both functional and probabilistic properties are preserved.
\footnote{If the program is a mathematical object, then  
as Andrew Vazonyi \cite{Vazsonyi:2002aa} pointed out: ``I'm not interested in \textit{ad hoc} solutions invented by clever people. I want a method that works for lots of problems\ldots\ One that mere mortals can use. Which is what a correctness-by-construction method should be.''}

\section{Enabling Correctness by Construction --- \pGCL}
\label{s1733}

The setting for correctness-by-construction of probabilistic programs is provided by \pGCL\ --the probabilistic Guarded-Command Language-- which contains both abstraction and (stepwise) refinement \cite{McIver:05a}. We begin by reviewing its origins, then its treatment of probabilistic choice and demonic choice, and finally its realisation of \CbC.

(This section can be skimmed on first reading: just collect \pGCL\ syntax from Figs.~\ref{f1200}--\ref{f1202}, and then skip directly to \Sec{s1736}.)

As we will not be treating non-terminating programs, we can base our description here on quite simple models for sequential (non-reactive) programs. The state space is some set $S$ and, in its simplest terms, a program takes an initial state to a final state: it (its semantics) therefore has type $S\,{\rightarrow}\,S$.
%\footnote{With non-termination we would instead have started with $S\,{\rightarrow}\,S\,{\cup}\,\{\bot\}$.}

The three subsections that follow describe logics based on successive enrichments of this, the simplest model,
and even the youngest of those logics is by now almost 25 years old: thus we will be ``reviewing'' rather than inventing.

The first enrichment, \Sec{s0910}, is based on the model $S\,{\rightarrow}\,\Pow S$ that allows demonic nondeterminism,\,%
\footnote{Constructor $\Pow$ is ``subsets of'' and $\Dist$ is ``discrete distributions on''.}
so facilitating abstraction; then in \Sec{s0911} the model $S\,{\rightarrow}\,\Dist S$ replaces demonic nondeterminism by probabilistic choice, losing abstraction (temporarily) but in its place gaining the ability to describe probabilistic outcomes; and finally in \Sec{s0912} the model $S\,{\rightarrow}\,\Pow\Dist S$ restores demonic nondeterminism, allowing programs that can abstract from precise probabilities. Using syntax we will make more precise in those sections, simple examples of the three increments in expressivity are
{\def\z#1{~~\makebox[7em][l]{#1}}
 \def\y#1{~~\parbox[t]{23em}{#1}}
\begin{enumerate}[(1)] \setlength\itemsep {0.5ex}
\item\label{i0939-1} \z{\PF[x:= H]} \y{Set variable \PF[x] to \PF[H] (as in any sequential language);}

\item\label{i0939-2} \z{\PF[x\In\ \{H,T\}]} \y{Set \PF[x]'s value demonically from the set $\{\PF[H],\PF[T]\}$;}

\item\label{i0939-3} \z{\PF[x\In\ H\,\PCNF{2}{3}\,T]} \y{Set \PF[x]'s value from the set $\{\PF[H],\PF[T]\}$ with probability \NF{2}{3} for \PF[H] and \NF{1}{3} for \PF[T], a ``biased coin''; and}

\item\label{i0939-4} \z{\PF[x\In\ H\,\PPC{\NF{1}{3}}{\NF{1}{3}}\,T]} \y{Set \PF[x] from the set $\{\PF[H],\PF[T]\}$ with probability \emph{at least} \NF{1}{3}\ each way, a ``capricious coin''.}
\end{enumerate}}

The last example of those \Itm{i0939-4} is the most general: for \Itm{i0939-3} is \PF[x\In\ H\,\PPC{\NF{2}{3}}{\NF{1}{3}}\,T]; and \Itm{i0939-2} is \PF[x\In\ H\,\PPC{0}{0}\,T]; and finally \Itm{i0939-1} is \PF[x\In\ H\,\PPC{1}{0}\,T]. 

\subsection{Floyd/Hoare/Dij\-kstra: pre- and postconditions: (\ref{i0939-1},\ref{i0939-2}) above}
\label{s0910}

We assume a typical sequential programming language with variables, expressions over those variables, assignment (of expressions to variables), sequential composition (semicolon or line break), conditionals and loops. It is more or less Dij\-kstra's \emph{guarded command language} \cite{Dijkstra:76}, and is based on the model $S\,{\rightarrow}\,\Pow S$, where $\Pow S$ is the set of all subsets of $S$.

\medskip
The \emph{weakest precondition} of program \Prog\ in such a language, with respect to a postcondition \Post\ given as a first-order formula over the program variables, is written \WPP{\Prog}{\Post} and means
\begin{quote}
the weakest formula (again on the program variables) that must hold \emph{before} \Prog\ executes in order to ensure that \Post\ holds \emph{after} \Prog\ executes \cite{Dijkstra:76}.
\end{quote}
In a typical compositional style, the \WP\ of a whole program is determined by the \WP\ of its components.

\begin{figure}\small
\begin{center}
\includegraphics[width=.6\textwidth]{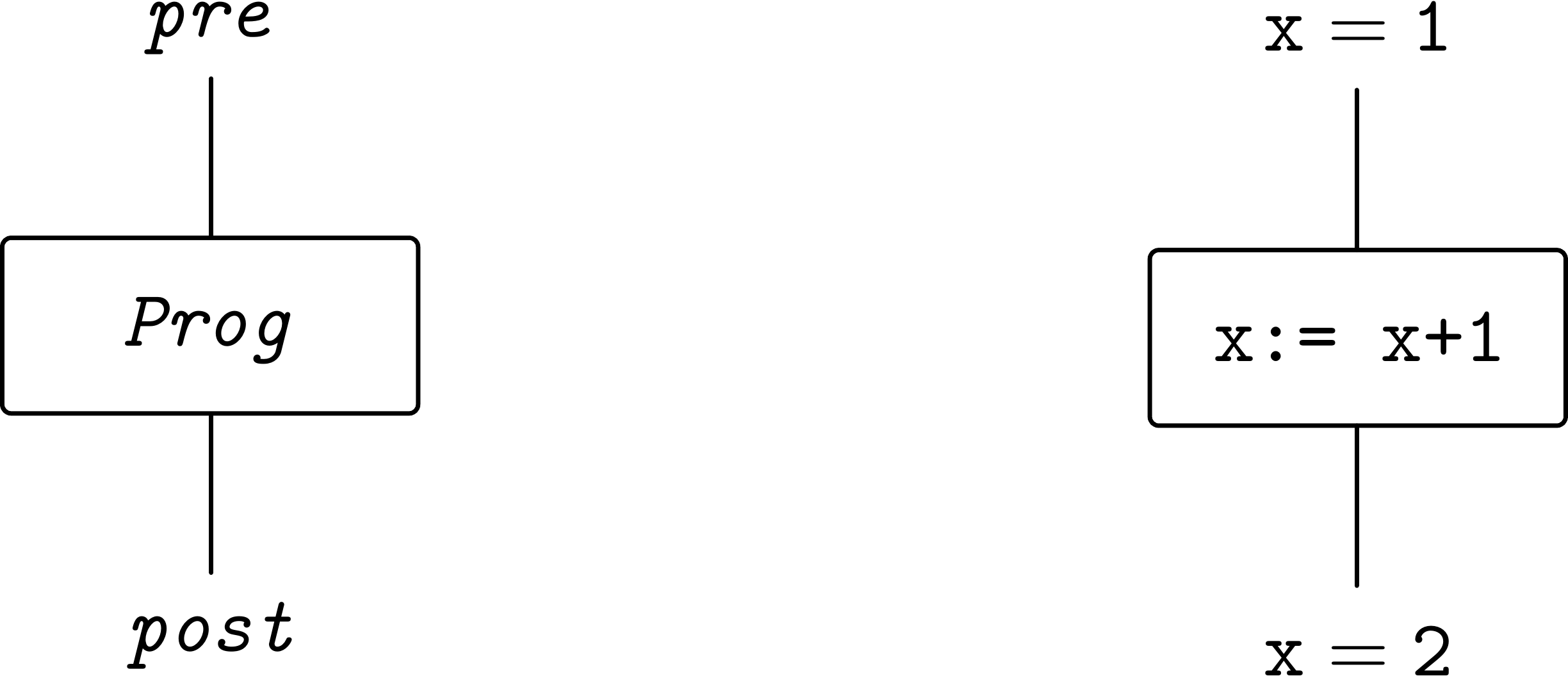}

\medskip
\begin{quote}
At left is a ``generic'' Floyd annotation of a flowchart containing only one program element. If the annotation \Meta{pre} holds ``on the way in'' to the program \Meta{Prog}, then annotation \Meta{post} will hold on the way out. At right is an example with specific annotations and a specific program.

\medskip
In the Hoare style the right-hand example would be written
\[\{{\tt x}=1\}~\ProgIL{x:= x+1}~\{{\tt x}=2\}\Qdot\]

In the Dijsktra style it would be written $\PF[x=1]\Implies \PF[\WPP{x:= x+1}{~x=2}]$.

They all three have the same meaning.
\end{quote}
\end{center}
\caption{Floyd-style annotated flowchart}\label{f1404}
\end{figure}

We group Dij\-kstra, Hoare and Floyd together because the Dij\-kstra-style implication $~\Pre\Implies\WPP{\Prog}{\Post}~$ has the same meaning as the Hoare-style triple $~\{\Pre\}~\Prog~\{\Post\}~$ which in turn has the same meaning as the original Floyd-style flowchart annotation, as  shown in \Fig{f1404} \cite{Floyd:67,Hoare:69}. All three mean ``If \Pre\ holds of the state before execution of \Prog, then \Post\ will hold afterwards.''

\medskip
Finally, a notable --but incidental-- feature of Dij\-kstra's approach was that (demonic) nondeterminism arose naturally, as an abstraction from possible concrete implementations.\,%
\footnote{See \Sec{step 5} for a further discussion of this.}
That is why we use $S\,{\rightarrow}\,\Pow S$ rather than $S\,{\rightarrow}\,S$ here. In later work (by others) that abstraction was made more explicit by including explicit syntax for a binary ``demonic choice'' between program fragments, a composition \Meta{\Left} \ND\ \Meta{\Right} that could behave either as the program \Left\ or as the program \Right. But that operator (\ND) was not really an extension of Dij\-kstra's work, because his (more verbose) conditional
\begin{ProgEqn}
IF\>\>True $\rightarrow$ \Left ~~~\Com If \PF[True] holds, then this branch may be taken.\\
\B\>\>True $\rightarrow$ \Right ~~\Com If \PF[True] holds, then also \emph{this} branch may be taken. \\
FI \Com (Dijkstra terminated all \PF[IF]'s with \PF[FI]'s.)~~~~~
\end{ProgEqn}%
was there in his original guarded-command language, introducing demonic choice naturally as an artefact of the program-design process --- and it expressed exactly the same thing. The (\ND) merely made it explicit.

\subsection{Kozen: probabilistic program logic: \Itm{i0939-3} above}
\label{s0911}
Kozen extended Dij\-kstra-style semantics to probabilistic programs, again over a sequential programming language but now based on the model $S\,{\rightarrow}\,\Dist S$, where $\Dist S$ is set of all discrete distributions in $S$.\,%
\footnote{Kozen's work did not restrict to discrete distributions; but that is all we need here.}
He replaced Dij\-kstra's demonic nondeterminism (\ND) by a ``probabilistic nondeterminism'' operator (\PC{p}) between programs, understood so that \ProgIL{\Left\,\PC{p}\;\Right} means ``Execute \Left\ with probability $p$ and \Right\ with probability $1{-}p$.'' The probability $p$ is (very) often \NF{1}{2} so that \ProgIL{coin:= Heads \PCF\ coin:= Tails} means ``Flip a fair coin.'' But probability \Meta{p} can more generally be any real number, and more generally still it can even be an expression in the program variables.

Kozen's corresponding extension of Floyd/Hoare/Dij\-kstra \cite{Kozen:81,Kozen:83} replaced Dij\-kstra's logical formulae with real-valued expressions (still over the program variables); we give examples below. The ``original'' Dij\-kstra-style formulae remain as a special case where real number 1 represents \True\, and 0 represents \False; and Dij\-kstra's definitions of \WP\ simply carry through essentially as they are\ldots\
except that an extra definition is necessary, for the new construct (\PC{\Meta{p}}), where Kozen defines that
\begin{align*}
	& \WPP{\Left\;\PC{\Meta{p}}\;\Right}{\Post} \\
	\textrm{is\qquad} &
	\Meta{p}\cdot\WPP{\Left}{\Post} ~+~ (1{-}\Meta{p})\cdot\WPP{\Right}{\Post}\Qdot
\end{align*}
With this single elegant extension, it turns out that in general \WPP{\Prog}{\Post} is the \emph{expected value}, given as a (real valued) expression over the \emph{initial} state, of what \Post\ will be in the \emph{final} state, i.e.\ after \Prog\ has finished executing from that initial state. (The initial/final emphasis simply reminds us that it is the same as for Dij\-kstra: the weakest precondition is what must be true in the \emph{initial} state for the postcondition to be true in the \emph{final} state.)
For example we have that
\begin{align*}
	\WPP{x:= 1-y\,\PCNF{1}{3}\;x:= 3*x}{~~~\PF[x]+3}
%	\WPP{{x:= 1-y\,\PCNF{1}{3}\;x:= 3\ast x}}{~~~\PF[x]+3}
	\WIDERM{is} \NF{1}{3}(1{-}\PF[y]+3) ~+~ \NF{2}{3}(3\PF[x]{+}3) 
	\quad,
\end{align*}
that is the real-valued expression $3\frac{1}{3} + 2\PF[x] - \PF[y]/3$ in which both \PF[x] and \PF[y] refer to their values in the initial state.\,%
%\footnote{Specific programs are written in \PF[this font], but with italics for meta-variables like \Left\ that stand for ``any program you like''. Formulae are written in ordinary mathematical notation, using however the program font for program variables within the formulae. Thus --a contrived example-- $(\exists k.~\PF[n]=2k)$ is a formula expressing that program variable \PF[n] is even.}

More impressive though is that if we introduce the convention that brackets $[-]$ convert Booleans to numbers, i.e.\ that $[\True] = 1$ and $[\False] = 0$, we have in general for \emph{Boolean}-valued \Prop\ the convenient idiom
\begin{align}
	& \WPP{\Prog}{[\,\Prop]} \label{e1520} \\
	\WIDERM{is} & \textrm{``the probability that \Prog\ establishes property \Prop''}
	,\,\footnotemark
	\nonumber
\end{align}
\footnotetext{The expected value of the characteristic function $[\,\Prop\,]$ of an event \Prop\ is equal to the probability that \Prop\ itself holds.}%
And if --further-- it happens that the ``probabilistic'' program \Prog\ actually contains no probabilistic choices at all, then \Eqn{e1520} just above has value 1 just when \Prog\ is guaranteed to establish \Post, and is 0 otherwise: it is in that sense that the Dij\-kstra-style semantics ``carries through'' into the Kozen extension. That is, if \Prog\ contains no probabilistic choice, and \Post\ is a conventional (Boolean valued) formula, then we have
\begin{align*}
	& \emph{Dij\-kstra style} & [\,\WPP{\Prog}{\Post}\,] \\ 
	\textrm{is the same as\qquad} & \emph{Kozen style} & \WPP{\Prog}{[\,\Post\,]}\HangRight{.\,\footnotemark}
\end{align*}
\footnotetext{Note that if \Prog\ contains (\PC{\Meta{p}}) somewhere, the above does not apply: Dij\-kstra semantics has no definition for (\PC{\Meta{p}}).}

The full power of the Kozen approach, however, starts to appear in examples like this one below:  we flip two fair coins and ask for the probability that they show the same face afterwards. Using the (Dij\-kstra) weakest-precondition rule that \WPP{\Meta{Prog1};\Meta{Prog2}}{~\Post} is simply \WPP{\Meta{Prog1}}{~\WPP{\Meta{Prog2}}{\Post}},\,%
\footnote{This is particularly compelling when \WP\ is Curried: sequential composition \WP(\Meta{Prog1};~\Meta{Prog2}) is then the functional composition $\WP(\Meta{Prog1})\circ\WP(\Meta{Prog2})$.}
we can calculate
\begin{Reason}
\Step{}{\WPP{{\PF c1:= H \PCF\, c1:= T;~c2:= H \PCF\, c2:= T}}{~~~[\PF[c1]=\PF[c2]]}}
\Step{=}{\WPP{{\PF c1:= H \PCF\, c1:= T}}{~~~\WPP{{\PF c2:= H \PCF\, c2:= T}}{[\PF[c1]=\PF[c2]]}}}
\Step{=}{\WPP{{\PF c1:= H \PCF\, c1:= T}}{~~~\NF{1}{2}[\PF[c1]=\PF[H]]+(1{-}\NF{1}{2})[\PF[c1]=\PF[T]]}}
\Space[-2ex]
\Step{=}{\NF{1}{2}(\NF{1}{2}[\PF[H]=\PF[H]]+\NF{1}{2}[H=\PF[T]]) + \NF{1}{2}(\NF{1}{2}[\PF[T]=\PF[H]]+\NF{1}{2}[\PF[T]=\PF[T]])
}
\Step{=}{\NF{1}{2}(\NF{1}{2}\cdot1+\NF{1}{2}\cdot0) + \NF{1}{2}(\NF{1}{2}\cdot0+\NF{1}{2}\cdot1)}
\Space[-2ex]
\Step{=}{\NF{1}{4}+\NF{1}{4}}
\Step{=}{\NF{1}{2}\textrm{~,\quad that is that the probability that $\PF[c1]\,{=}\,\PF[c2]$ is $\NF{1}{2}$.}}
\end{Reason}
A nice further exercise for seeing this probabilistic \WP\ at work is to repeat the above calculation when one of the coins uses (\PC{\Meta{p}}) but (\PCF) is retained for the other, confirming that the answer is still \NF{1}{2}.

\begin{figure}\small
\def\z{13em}
\def\y{17em}
\def\Pa{\Meta{Prog1}}
\def\Pb{\Meta{Prog2}}
\begin{center}\begin{tabular}{p{6em}@{~~}ll}
\underline{name} & \underline{syntax} & \underline{semantics} \\\\
expectation \Post& \parbox[t]{\z}{real-valued expression \\ over the program variables} & (the usual) \\[3.5ex]
expression \Exp& \parbox[t]{\z}{expression over the program \\ variables (of any type)} & (the usual) \\[3.5ex]
condition \Cond& \parbox[t]{\z}{Boolean-valued expression \\ over the program variables} & (the usual)
\\[3.5ex]
substitution & \Sbst{\Meta{E1}}{\Meta{x}}{\Meta{E2}} & \parbox[t]{\y}{Replace all free occurrences of \Meta{x} in \Meta{E1} by \Meta{E2} (with the usual caveats.)} \\
\\\hline
\\
assignment & \PF[\Meta{x}:= \Exp] & Evaluate \Exp; assign it to \Meta{x}. \\
&& \WPP{\Meta{x}:= \Exp}{~\Post} = \Sbst{\Post}{\Meta{x}}{\Exp}\\
\\
sequential composition & \Pa;\Pb & Execute \Pa\ then \Pb. \\[.5ex]
&\multicolumn{2}{r}{\WPP{\Pa;\Pb}{~\Post} = \WPP{\Pa}{~\WPP{\Pb}{\Post}}}
\\[3.5ex]
conditional & \PF IF \Cond~THEN \Pa\ ELSE \Pb
& ~~~\parbox[t]{\y}{Evaluate Boolean \Cond, then execute \\\Pa\ or \Pb\ accordingly.} \\\\[-1ex]
&\multicolumn{2}{r}{\WPP{\PF IF \Cond~THEN \Pa\ ELSE \Pb}{~\Post}} \hspace{.4em}\\
&\multicolumn{2}{r}{=\quad$[\Cond]{\cdot}\WPP{\Pa}{\Post}+[\neg\Cond]{\cdot}\WPP{\Pb}{\Post}$}
\\[3.5ex]
loop & \PF WHILE \Cond~DO \Prog
& \parbox[t]{\y}{Evaluate Boolean \Cond, then execute \Prog\ (and repeat), or exit, accordingly.} \\\\[-1ex]
\multicolumn{3}{r}{\quad The usual least fixed point, based on} \\
\multicolumn{3}{r}{\quad \(\PF[WHILE \Cond~DO \Prog] \quad=\quad
                           \PF[IF \Cond~THEN (\Prog;~WHILE \Cond~DO \Prog)]
                         \)} \\
\end{tabular}\end{center}

\bigskip
The above cases cover the constructs of \pGCL\ without probabilistic- or demonic choice, but nevertheless defined with Kozen-style ``numeric'' \WP's which, applied to ``post-expectations'' give ``pre-expectations''.
\caption{Syntax and \WP-semantics for ``restricted'' \pGCL}\label{f1200}
\end{figure}

\subsection{McIver/Morgan: pre- and post-expectations}
\label{s0912}

Following Kozen's probabilistic semantics at \Sec{s0911} just above (which itself turned out later to be a special case of Jones and Plotkin's probabilistic powerdomain contruction \cite{Jones:89}) we restored demonic choice to the programming language and called it \pGCL\ \cite{Morgan:96d,McIver:05a}. It contains both demonic (\ND) and probabilistic (\PC{\Meta{p}}) choices; its model is $S\,{\rightarrow}\,\Pow\Dist S$; and it is the language we will use for the correct-by-construction program development we carry out below \cite{McIver:05a}. Figures \ref{f1200}--\ref{f1202} summarise its syntax and its \WP-logic.

To illustrate demonic- vs.\ probabilistic choice, we'll revisit the two-coin program from above. This time, one coin will have a probability-$p$ bias for some constant $0\,{\leq}\,p\,{\leq}\,1$ (thus acting as a fair coin just when $p$ is \NF{1}{2}). The other choice will be purely demonic.

\begin{figure}\small
\def\z{10em}
\def\y{18em}
\def\Pa{\Meta{Prog1}}
\def\Pb{\Meta{Prog2}}
\begin{center}\begin{tabular}{p{6em}l@{~~~}l}
\underline{name} & \underline{syntax} & \underline{semantics} \\\\
probabilistic choice & \Pa\,\PC{\Meta{p}} \Pb & \parbox[t]{\y}{Evaluate \Meta{p}, which must be in $[0,1]$, then execute \Pa\ with that probability; otherwise execute \Pb.}\\\\[-1ex]
\multicolumn{3}{r}{\WPP{\Pa \PC{\Meta{p}} \Pb}{~\Post} ~$=$~ \Meta{p}$\cdot$\WPP{\Pa\!}{\,\Post}+(1{-}\Meta{p})$\cdot$\WPP{\Pb\!}{\,\Post}} \\
\\\\
demonic choice & \Pa $\;\ND$ \Pb & \parbox[t]{\y}{Choose demonically whether to execute \Pa\ or \Pb.}\\\\[-1ex]
\multicolumn{3}{r}{\WPP{\Pa\,\ND\;\Pb}{~\Post} ~$=$~ \WPP{\Pa\!}{\,\Post}~\Min~\WPP{\Pb\!}{\,\Post}} \\
\\\\
\end{tabular}\end{center}
\vspace{-2em}
These ``extra'' cases cover the probabilistic- and demonic choice constructs of \pGCL.
\caption{Syntax and \WP-semantics for \pGCL's choice constructs}\label{f1201}
\end{figure}

\begin{figure}\small
\def\z{10em}
\def\y{14em}
\def\Pa{\Meta{Prog1}}
\def\Pb{\Meta{Prog2}}
\begin{center}\begin{tabular}{p{6em}ll}
\underline{name} & \underline{syntax} & \underline{semantics} \\\\

do nothing & \parbox{\z}{\PF[SKIP]} & \parbox{\y}{\WPP{\PF[SKIP]}{\,\Post} = \Post~.} \\
fail & \PF[ABORT] & \WPP{\PF[ABORT]}{\,\Post} = 0~. \\\\
probabilistic assignment & \PF[\Meta{x}\In\ \Meta{E1}\PC{\Meta{p}}\,\Meta{E2}] & As for \PF[(\Meta{x}:= \Meta{E1})\;\PC{\Meta{p}} (\Meta{x}:= \Meta{E2})]~. \\
\\
demonic assignment & \PF[\Meta{x}\In\ \Meta{E1}\,\ND\;\Meta{E2}] & As for \PF[(\Meta{x}:= \Meta{E1}) \ND\ (\Meta{x}:= \Meta{E2})]~. \\
\\
probabilistic & \PF IF~\Meta{p}~THEN \Pa & As for \PF[\Pa\,\PC{\Meta{p}}\;\Pb]~.\\
conditional   &\PF~~~~~\,ELSE \Pb
\\\\
probabilistic\qquad  & \PF WHILE \Meta{p}~DO \Prog & As for ordinary loop,\\
loop   && but using probabilistic conditional.
\end{tabular}\end{center}

\bigskip
The cases above introduce special commands, abbreviations and ``syntactic sugar'' for \pGCL.

\medskip
Command \PF[SKIP] allows an ``\PF[ELSE]-less'' conditional, as used e.g.\ in \Fig{f1200}, to be defined in the usual way, as \PF[IF \Cond~THEN \Pa\ ELSE SKIP].

\medskip
Command \PF[ABORT] allows \WPP{\PF[WHILE \Cond~DO \Prog]}{~\Post}, as a least fixed point, to be defined as the supremum of
\begin{quote}\begin{tabular}{l}
\WPP{\PF[ABORT]}{~\Post} \\
\WPP{\PF[IF \Cond~THEN (\Prog;ABORT)]}{~\Post} \\
\WPP{\PF[IF \Cond~THEN (\Prog;(IF \Cond~THEN (\Prog;ABORT)))]}{~\Post}\\
$\vdots$\quad,
\end{tabular}\end{quote}
which exists (in spite of the reals' being unbounded) because it can be shown by structural induction that
\[
 \WPP{\Prog}{\Post} \Wide{\leq} \Post \textrm{\quad,}
\]
and that \WPP{\Prog}{$-$} is continuous, for all programs \Prog. The above is therefore a chain, is dominated by \Post\ itself, and attains the limit at $\omega$.

\caption{Syntax and \WP-semantics for \pGCL's choice constructs}\label{f1202}
\end{figure}

We start with the (two-statement) program
%\[
%\begin{minipage}{\textwidth}xxx\end{minipage}
%\]
\begin{ProgEqn}
	c1:= H \PC{p}\, c1:= H \\ c2:= H \;\,\ND\, c2:= T \textrm{\quad,}
\end{ProgEqn}%
where the first statement is probabilistic and the second is demonic,
and ask, as earlier, ``What is the probability that the two coins end up equal?''
%(Sequential composition is indicated by the line-break.)
We calculate
\begin{Reason}
\Step{}{\WPP{{\PF c1:= H \PC{p}\, c1:= T;~c2:= H {\ND}\;c2:= T}}{~~~[\PF[c1]=\PF[c2]]}}
\Step{=}{\WPP{{\PF c1:= H \PC{p}\, c1:= T}}{~~~\WPP{{\PF c2:= H {\ND}\;c2:= T}}{~[\PF[c1]=\PF[c2]]}}}
\Step{=}{\WPP{{\PF c1:= H \PC{p}\, c1:= T}}{~~~[\PF[c1]=\PF[H]] ~\Min~ [\PF[c1]=\PF[T]]}}
\Space[-2ex]
\Step{=}{p\,{\cdot}([\PF[H]=\PF[H]] ~\Min~ [\PF[H]=\PF[T]]) + (1{-}{\it p}){\cdot}([\PF[T]=\PF[H]] ~\Min~ [\PF[T]=\PF[T]])}
\Step{=}{p\,{\cdot}(1 ~\Min~ 0) + (1{-}p){\cdot}(0 ~\Min~ 1)}
\Step{=}{p\,{\cdot}0 + (1{-}p){\cdot}0}
\Space[-2ex]
\Step{=}{0\quad,}
\end{Reason}
to reach the conclusion that the probability of the two coins' being equal finally\ldots\ is zero. And that highlights the way demonic choice is usually treated: it's a worst-case outcome. The ``demon'' --thought of as an agent-- always tries to make the outcome as bad as possible: here because our desired outcome is that the coins be equal, the demon always sets the coin \PF[c2] so they will differ. If we repeated the above calculation with postcondition $\PF[c1]{\neq}\PF[c2]$ instead, the result would \emph{again} be zero: if we change our minds, want the coins to differ, then the demon will change his mind too, and act to make them the same.\,%
\footnote{This is not a novelty: demonic choice is usually treated that way in semantics --- that's why it's called ``demonic''.}

Implicit in the above treatment is that the \PF[c2] demon knows the outcome of the \PF[c1] flip --- which is reasonable because that flip has already happened by the time it's the demon's turn.

Now we reverse the statements, so that the demon goes first: it must set \PF[c2] without knowing beforehand what \PF[c1] will be. The program becomes

\vspace{-3ex}
\begin{ProgEqn}
	c2:= H \;\,\ND\, c2:= T \\
	c1:= H \,\PC{p} c1:= T \textrm{\quad,}
\end{ProgEqn}%
and we calculate
\begin{Reason}
\Step{}{\WPP{{\PF c2:= H \,\ND\;c2:= T;~c1:= H \PC{\it p}\, c1:= T}}{~~~[\PF[c1]=\PF[c2]]}}
\Step{=}{\WPP{{\PF c2:= H \,\ND\;c2:= T}}{~~~\WPP{{\PF c1:= H \PC{\it p}\, c1:= T}}{~[\PF[c1]=\PF[c2]]}}}
\Step{=}{\WPP{{\PF c2:= H \,\ND\;c2:= T}}{~~~p\,{\cdot}[\PF[H]=\PF[c2]] + (1{-}p){\cdot}[\PF[T]=\PF[c2]]}}
\Step{=}{p\,{\cdot}[\PF[H]=\PF[H]] + (1{-}p){\cdot}[\PF[T]=\PF[H]]~~\Min~~p\,{\cdot}[\PF[H]=\PF[T]] + (1{-}p){\cdot}[\PF[T]=\PF[T]]}
\Step{=}{p\,{\cdot}1 + (1{-}p){\cdot}0~~\Min~~p\,{\cdot}0 + (1{-}p){\cdot}1}
\Step{=}{p~\Min~(1{-}p)\Qdot}
\end{Reason} 
Since the demon set flip \PF[c2] \emph{without} knowing what the \PF[c1]-flip would be (because it had not happened yet), the worst it can do is to choose \PF[c2] to be the value that it is known \PF[c1] is least likely to be --- which is just the result above, the lesser of $p$ and $1{-}p$. If --as before-- we change our minds and decide instead that we would like the coins to be different, then the demon adapts by choosing \PF[c2] to be the value that \PF[c1] is \emph{most} likely to be.

Either way, the probability our postcondition will be achieved, the pre-expectation of its characteristic function, is the same $p\,\Min\,(1{-}p)$ --- so that only when $p\,{=}\,\FF$, i.e.\ when $p\,{=}\,(1{-}p)$, does the demon gain no advantage.

\section{Probabilistic \emph{correctness by construction} in action\,\protect\footnotemark}\label{s1736}
\footnotetext{This intent of this section can be understood based on the syntax given in Figs.~\ref{f1200}--\ref{f1202}.}

Our first example problem conceptually will be to achieve a binary choice of arbitrary bias using only a fair coin. With the apparatus of \Sec{s0912} however, we can immediately move from conception to precision:
\begin{quote}
We must write a \pGCL\ program that implements \ProgIL{\Left\,\PC{p}\;\Right},\linebreak
under the constraint that the only probabilistic choice operator we are allowed to use in the final (\pGCL) program is $(\PCF)$.
\end{quote}
This is not a hard problem mathematically: the probabilistic calculation that solves it is elementary. Our point here is to use this simple problem to show how such solutions can be calculated within a programming-language context, while maintaining rigour (possibly machine-checkable) at every step.

The final program is given at \Eqn{p0955} in \Sec{step 5}.

\subsection{Step 1 --- a simplification}\label{step 1}
We'll start by simplifying the problem slightly, instantiating the programs \Left\ and \Right\ to \PF[x:= 1] and \PF[x:= 0] respectively. Our goal is thus to implement
\begin{ProgEqn}[p1643]
	x\In\ 1\,\PC{p}\,0\Qcomma
\end{ProgEqn}%
for arbitrary $p$, and our first step is to create two other distributions $1\,{\PC{q}}\,0$ and $1\,{\PC{r}}\,0$ whose average is $1\,{\PC{p}}\,0$ --- that is
\begin{equation}\label{e1635}
	\FF\times(\,(1\,{\PC{q}}\,0)+(1\,{\PC{r}}\,0)\,) \Wide{=} (1\,{\PC{p}}\,0)\Qdot
\end{equation}
A fair coin will then decide whether to carry on with $1\,\PC{q}\,0$ or with $1\,\PC{r}\,0$.

Trivially \Eqn{e1635} holds just when $(q{+}r)/2 = p$, and if we represent $p,q,r$ as variables in our program, we can achieve \Eqn{e1635} by the double assignment
\begin{ProgEqn}[p1720]
	IF p\,$\leq$\,\FF\ $\rightarrow$ q,r:= 0,2p \\
	\B\ p\,$\geq$\,\FF\ $\rightarrow$ q,r:= 2p-1,1 \\
	\HangLeft{\footnotemark\qquad}%
	FI \\
	\Ass{p = (q+r)/2}\Qcomma
\end{ProgEqn}%
\footnotetext{We will sometimes include Dij\-kstra's closing \PF[FI].}%
whose postcondition indicates what the assignment has established. If we follow that with a fair-coin flip between continuing with \PF[q] or with \PF[r], viz.
\begin{ProgEqn}[p1714]
	IF p\,$\leq$\,\FF\ $\rightarrow$ q,r:= 0,2p \Com Here \PF[q] is 0.\\
	\B\ p\,$\geq$\,\FF\ $\rightarrow$ q,r:= 2p-1,1 \Com Here \PF[r] is 1. \\
	FI \\
%	\Ass{p = (q+r)/2} \\
	(x\In\ \OZ{q}) \PCF\ (x\In\ \OZ{r}) \Com The fair coin $(\PCF)$ here is permitted.
\end{ProgEqn}%
then we should have implemented \Prg{p1643}. But what have we gained? 

The gain is that, whichever branch of the conditional is taken, there is a \FF\ probability that the problem we have \emph{yet} to solve will be either $(\PC{0})$ or $(\PC{1})$, both of which are trivial. If we were unlucky, well\ldots\ then we just try again.
But how do we show rigorously that \Prg{p1643} and \Prg{p1714} are equal?

If we look back at \Prg{p1720}, we find the assertion \Ass{p = (q+r)/2} which is easy to establish by conventional Hoare-logic or Dij\-kstra-\WP\ reasoning from the conditional just before it. (We removed it from \Prg{p1714} just to reduce clutter.) Rigour is achieved by calculating

%\begin{ProgEqn}
%	IF p\,$\geq$\,\FF\ $\rightarrow$ q,r:= 2p-1,1 \\
%	\B\ p\,$\leq$\,\FF\ $\rightarrow$ q,r:= 0,2p \\
%	FI \\
%	\Ass{p = (q+r)/2} \\
%	(x\In\ \OZ{q}) \PCF\ (x\In\ \OZ{r})
%\end{ProgEqn}%

%\begin{Reason}
%\Step{}{\WPP{(x\In\ \OZ{q})\PCF\,(x\In\ \OZ{r})}{\,[\PF[x]=1]}}
%\Space
%\Step{$=$}{\FF\;\WPP{(x\In\ \OZ{q})}{{\,[\PF[x]=1]}}
%           ~+~ \FF\;\WPP{(x\In\ \OZ{r})}{{\,[\PF[x]=1]}}}
%\Space
%\Step{$=$}{\NF{\PF q}{2}[1=1] + \NF{(1-\PF[q])}{2}[0=1] 
%           + \NF{\PF r}{2}[1=1] + \NF{(1-\PF[r])}{2}[0=1]}
%\Space
%\Step{$=$}{(\PF[q]+\PF[r])/2}
%\StepR{$=$}{\Ass{p = (q+r)/2}}{\PF[p]}
%\Step{$=$}{\WPP{x\In\ \OZ{p}}{\,[\PF[x]=1]}\Qdot}
%\end{Reason}
%
%For any post-expectation,

\begin{Reason}
\Step{}{\WPP{(x\In\ \OZ{q})~\PCF~(x\In\ \OZ{r})}{~~~\Post}}
\Space[-2ex]
\Step{$=$}{\FF\;\WPP{(x\In\ \OZ{q})}{\Post}
           ~+~ \FF\;\WPP{(x\In\ \OZ{r})}{\Post}}
\Step{$=$}{\NF{\PF q}{2}\cdot\Sbst{\Post}{\PF[x]}{1} + \NF{(1-\PF[q])}{2}\cdot\Sbst{\Post}{\PF[x]}{0} 
           + \NF{\PF r}{2}\cdot\Sbst{\Post}{\PF[x]}{1} + \NF{(1-\PF[r])}{2}\cdot\Sbst{\Post}{\PF[x]}{0}}
\Step{$=$}{(\PF[q]{+}\PF[r])/2\cdot\Sbst{\Post}{\PF[x]}{1} ~+~ (1-(\PF[q]{+}\PF[r])/2)\cdot\Sbst{\Post}{\PF[x]}{0}}
\StepR{$=$}{\Ass{p = (q+r)/2}}{\PF[p]\cdot\Sbst{\Post}{\PF[x]}{1} ~+~ (1{-}\PF[p])\cdot\Sbst{\Post}{\PF[x]}{0}}
\Space[-2ex]
\Step{$=$}{\WPP{x\In\ \OZ{p}}{~~~\Post}\Qcomma}
\end{Reason}%
for arbitrary postcondition \Post\ where at the end we used \Ass{p = (q+r)/2}. Thus $\Eqn{p1643}\,{=}\, \Eqn{p1714}$ because for any \Post\ their pre-expectations agree.

\subsection{Step 2 --- intuition suggests a loop}

We now return to the remark ``\ldots\ then we just try again.'' If we replace the final fair-coin flip \ProgIL{(x\In\ \OZ{q}) \PCF\ (x\In\ \OZ{r})} by \ProgIL{p\In\ q\,{\PCF}\,r} then \mbox{--intuitively--} we are in a position to ``try again'' with \ProgIL{x\In\ 1\,{\PC{p}}\,0}. Although it is the same as the statement we started with, we have made progress because variable \PF[p] has been updated --- and with probability \FF\ it is either 0 or 1 and we are done. If it is not, then we arrange for a second execution of
\begin{ProgEqn}[p1852]
	IF  p\,$\leq$\,\FF\ $\rightarrow$ q,r:= 0,2p \\
	\B\ p\,$\geq$\,\FF\ $\rightarrow$ q,r:= 2p-1,1 \\
	FI \\
	p\In\ q\,{\PCF}\,r
\end{ProgEqn}%
and, if \emph{still} \PF[p] is neither 0 nor 1, then \ldots\ we need a loop.

\subsection{Step 3 --- introduce a loop}\label{step 3}
We have already shown that
\[
	\Prg{p1643} \Wide{=} \Prg{p1852};~\Prg{p1643}\Qdot
\]
A general equality for sequential programs (including probabilistic) tells us that in that case also we have
\[
	\Prg{p1643} \Wide{=} \PF[WHILE \Cond\ DO $\Prg{p1852}$ OD; $\Prg{p1643}$]\HangRight{\quad\footnotemark}
\]
\footnotetext{As before, we usually use Dij\-kstra's loop-closing \PF[OD]\,.}%
for any loop condition \Cond, provided the loop terminates. Intuitively that is clear because, if \Prg{p1643} can annihilate \Prg{p1852} once from the right, then it can do so any number of times. A rigorous argument appeals to the fixed-point definition of \PF[WHILE], which is where termination is used. (If \Cond\ were \PF[False], so that the loop did not terminate, the \RHS\ would be \PF[Abort], thus providing a clear counter-example.)

For probabilistic loops, the usual ``certain''  termination is replaced with \emph{almost-sure} termination, abbreviated \AST, which means that the loop terminates with probability one: put the other way, that would be that the probability of iterating forever is zero. For example the program
\begin{ProgEqn}
	c:= H; WHILE c=H DO c\In\ H\,\PCF\;T OD\Qdot
\end{ProgEqn}
terminates almost surely because the probability of flipping \PF[T] forever is zero.

A reasonably good \AST\ rule for probabilistic loops is that the variant is (as usual) a natural number, but must be bounded above; and instead of having to decrease on every iteration, it is sufficient to have a non-zero probability of doing so \cite{Morgan:96b,McIver:05a}.\,%
\footnote{By ``reasonably good'' we mean that it deals with most loops, but not all: it is sound, but not complete. There are more complex rules for dealing with more complex situations \cite{McIver:2017aa}. Strictly speaking, over infinite state spaces ``non-zero'' must be strengthened to ``bounded away from zero'' \cite{Morgan:96b}.}
The variant for our example loop just above is {\PF{}[c=H]}, which has probability \FF\ of decreasing from {\PF{}[H=H]}, that is 1, to {\PF{}[T=H]} on each iteration.

The loop condition \Cond\ for our program will be $0\,{<}\,\PF[p]\,{<}\,1$ and the variant comes directly from there: it is {\PF{}[0$<$\PF[p]$<$1]}, which has probability of \FF\ of decreasing from 1 to 0 on each iteration: and when it is 0, that is $0\,{<}\,\PF[p]\,{<}\,1$ is false, the loop must exit. With that, we have established that our original \Prg{p1643} equals the looping program
\begin{ProgEqn}
	WHILE 0\,<\,p\,<\,1 DO \+\\
		IF  p\,$\leq$\,\FF\ $\rightarrow$ q,r:= 0,2p \\
		\B\ p\,$\geq$\,\FF\ $\rightarrow$ q,r:= 2p-1,1 \\
		FI \\
		p\In\ q\,{\PCF}\,r \-\\
	OD \\
	\Ass{\PF[p]=1 \lor \PF[p]=0} \\
	x\In\ 1\,{\PC{\PF[p]}}\,0\Qcomma
\end{ProgEqn}%
where the assertion at the loop's end is the negation of the loop guard.

\subsection{Step 4 --- use the loop's postcondition}

There is still the final \PF[x\In\ 1\,{\PC{\PF[p]}}\,0] to be dealt with, at the end; but the assertion \Ass{\PF[p]=1 \lor \PF[p]=0} just before it  means that it executes only when \PF[p] is zero or one. So it can be replaced by \PF[IF p=0 THEN x\In\ 1\PC{1}0 ELSE x\In\ 1\PC{0}0]~,
%\begin{ProgEqn} IF p=0 THEN x\In\ 1\PC{1}0 ELSE x\In\ 1\PC{0}0\Qcomma\end{ProgEqn}%
i.e.\ with just \ProgIL{x:= p}. Mathematically, that would be checked by showing for all post-expectations \Post\ that
\[
	\PF[p]=1 \lor \PF[p]=0 \Wide{\Implies}\WPP{x\In\ 1\,{\PC{\PF[p]}}\,0}{\Post} = \WPP{x:= p}{\Post}\Qdot
%	\begin{array}[t]{cl} &\WPP{x\In\ 1\,{\PC{\PF[p]}}\,0}{\Post} \\
%		= & \WPP{x:= p}{\Post}\Qdot
%	\end{array}
\]
But it's a simple-enough step just to believe (unless you were using mechanical assistance, in which case it \emph{would} be checked).

And so now the program is complete: we have implemented \ProgIL{x\In\ 1\PC{p}0} by a step-by-step correctness-by-construction process that delivers the program
\begin{ProgEqn}[p0925]
	WHILE 0\,<\,p\,<\,1 DO \+\\
		IF  p\,$\leq$\,\FF\ $\rightarrow$ q,r:= 0,2p \\
		\B\ p\,$\geq$\,\FF\ $\rightarrow$ q,r:= 2p-1,1 \\
		FI \\
		p\In\ q\,{\PCF}\,r \-\\
	OD \\
	x:= p
\end{ProgEqn}
in which only fair choices appear. And each step is provably correct.

\subsection{Step 5 --- after-the-fact optimisation}\label{step 5}
There is still one more thing that can (provably) be done with this program, and it's typical of this process: only when the pieces are finally brought together do you notice a further opportunity. It makes little difference --- but it is irresistible.

Before carrying it out, however, we should be reminded of the way in which these five steps are isolated from each other, how all the layers are independent. This is an essential part of \CbC, that the reasoning can be carried out in small, localised areas, and that it does not matter --for correctness-- where the reasoning's target came from; nor does it matter where it is going.

Thus even if we had absolutely no idea what \Prg{p0925} was supposed to be doing, still we would be able to see that if we are replacing \PF[x] by \PF[p] at the end, we could just as easily replace it at the beginning; and then we can remove the variable \PF[p] altogether. That gives
\begin{ProgEqn}[p0955]
	\LCom Now $p$ is again a parameter, as it was in the original specification. \\
	x:= $p$ \\
	WHILE 0\,<\,x\,<\,1 DO \+\\
		IF  x\,$\leq$\,\FF\ $\rightarrow$ q,r:= 0,2x \Com When $\PF[x]\,{=}\,\FF$, these two \\
		\B\ x\,$\geq$\,\FF\ $\rightarrow$ q,r:= 2x-1,1 \Com branches have the same effect. \\
		FI \\
		x\In\ q\,{\PCF}\,r \-\\
	OD\Qcomma \\
	\LCom The above implements \ProgIL{x\In\ 1\PC{p}0} for any $0\,{\leq}\,p\,{\leq}\,1$.
\end{ProgEqn}%
and we are done. When $p$ is 0 or 1, it takes no flips at all; when $p$ is \FF, it takes exactly one flip; and for all other values the expected number of flips is 2.

We notice that \Prg{p0955} appears to contain demonic choice, in that when $\PF[x]\,{=}\,\FF$ the conditional could take either branch. The nondeterminism is real --- even though the \emph{effect} is the same in either case, that \ProgIL{q,r:= 0,1} occurs. But genuinely different computations are carried out to get there: in the first branch $2(\FF)\,-\,1$ is evaluated to 0; and in the second branch $2(\FF)$ is evaluated to 1.

This is not an accident: we recall from \Sec{s0910} that for Dij\-kstra such nondeterminism arises naturally
%\Footnote{Get the exact quote.}
as part of the program-construction process. Where did it come from in this case?

The specification from which this conditional \PF[IF $\cdots$ FI] arose was set out much earlier, at \Eqn{e1635} which given $p$ has many possible solutions in $q,r$. One of them for example is $q\,{=}\,r\,{=}\,p$ which however would have later given a loop whose non-termination would prevent Step 3 at \Sec{step 3}. With an eye on loop termination, therefore, we took a design decision that at least one of $q,r$ should be ``extreme'', that is 0 or 1. To end up with $q\,{=}\,0$, what is the largest that $p$ could be without sending $r$ out of range, that is strictly more than 1? It's $p\,{=}\,\FF$, and so the first \PF[IF]-condition is $p\,{\leq}\,\FF$.\,%
The other condition $p\,{\geq}\,\FF$ arises similarly, and it absolutely does not matter that they overlap: the program will be correct whichever \PF[IF]-branch taken in that case.

And, in the end --in \Eqn{p0955} just above-- we see that indeed that is so.

\section{Implementing \emph{any} discrete choice with a fair coin}\label{s1041}

Suppose instead of trying to implement a biased coin (as we have been doing so far), we want to implement a general (discrete) probabilistic choice of \PF[x]'s value from its type, say a finite set \CalX, but still using only a fair coin in the implementation. An example would be choosing \PF[x] uniformly from $\{x_0,x_1,x_2\}$, i.e.\ a three-way fair choice. But what we develop below will work for any discrete distribution on a finite set \CalX\ of values: it does not have to be uniform.

The combination of probability \emph{and} abstraction allows a development like the one in \Sec{s1736} just above to be replayed, but a greater level of generality. We begin with a variable \PF[d] of type $\Dist\CalX$,\,%
\footnote{Recall from \Sec{s0911} that $\Dist\CalX$ is the set of discrete distributions over finite set \CalX.}
where we recall that \CalX\ is the type of \PF[x]; and our specification is \ProgIL{x\In\ d}, that is ``Set \PF[x] according to distribution \PF[d].''

%The specification is \ProgIL{x\In\ d}, where \PF[x] is of some type \PF[T] and \PF[d] is of type \PF[\Dist\,T], that is distributions on \PF[T]. Assume for now that \PF[T] is finite, and \PF[d] is (therefore) discrete. The implementation must do the same thing, but use only (\PCF).

\begin{figure}\small
\def\z{10em}
\def\Pa{\Meta{Prog1}}
\def\Pb{\Meta{Prog2}}
\begin{center}\begin{tabular}{p{10em}ll}
\underline{name} & \underline{syntax} & \underline{semantics} \\\\
choose from set & \PF[\Meta{x}\In\ \Meta{set}] \\[0.5ex]
\multicolumn{3}{r}{\hspace{\z} \WPP{\PF[\Meta{x}\In\ \Meta{set}]}{~\Post} = (\Min\,$e\,{\mid}\,e\,{\in}\,\Meta{set}$\,. \Sbst{\Post}{\Meta{x}}{$e$})} \\\\
assign ``such that'' & \PF[\Meta{x}\St\ \Meta{property}(\Meta{x})] \\[0.5ex]
\multicolumn{3}{r}\qquad \WPP{\PF[\Meta{x}\St\ \Meta{property}(\Meta{x})]}{~\Post} = (\Min\,$e\,{\mid}$\,\Meta{property}($e$)\,. \Sbst{\Post}{\Meta{x}}{$e$}) \\\\
\end{tabular}\end{center}

\bigskip
The above generalise to more than a single variable, and are consistent with the earlier definitions: thus
\begin{Reason}
\Step{}{\ProgIL{x:= a\;\,\ND\,\;x:= b}}
\Step{$=$}{\ProgIL{x\In\ \{a,b\}}}
\Step{$=$}{\ProgIL{x\St\ x$\,\in\,$\{a,b\}}\Qdot}
\end{Reason}

\bigskip
By analogy with ``choose from set'' (but not itself an abstraction) we have also

\medskip
\begin{center}\begin{tabular}{p{15em}ll}
\underline{name} & \underline{syntax} & \underline{semantics} \\\\
choose from distribution & \PF[\Meta{x}\In\ \Meta{dist}] \\[0.5ex]
\multicolumn{3}{r}{\hspace{3em}\WPP{\PF[\Meta{x}\In\ \Meta{dist}]}{~\Post} = ($\sum\,e\,{\mid}\,e\,{\in}\,\Supp{\Meta{dist}}$\,. \Meta{dist}($e$)$\,\cdot\,$\Sbst{\Post}{\Meta{x}}{$e$})} \Qcomma\\\\
\end{tabular}\end{center}
where $\Meta{dist}(e)$ is the probability that \Meta{dist} assigns to $e$ and \Supp{\Meta{dist}} is the \emph{support} of \Meta{dist}, the set of elements to which it assigns non-zero probability.\,\footnotemark%

\medskip
It is just the expected value of \Post, considered as a function of \Meta{x}, over the distribution \Meta{dist} on \Meta{x}. (Since \ProgIL{E1\,\PC{p}\,E2} is a distribution, the definition above agrees with the earlier meaning of \ProgIL{x\In\ E1\,\PC{p}\,E2} that we gave in \Fig{f1202} as an abbreviation.)

\caption{Abstraction in \pGCL}\label{f1314}
\end{figure}
\footnotetext{Summing over all possible values $e$ of \Meta{x} would give the same result, since the extra values have probability zero anyway. Some find this formulation more intuitive.}

\subsection{Replaying earlier steps from \Sec{s1736}}
Our first step --Step 1-- is to declare two more $\Dist\CalX$-typed variables \PF[d0] and \PF[d1], and --as in \Sec{step 1}-- specify that they must be chosen so that their average is the original distribution \PF[d]; for that we use the \pGCL\ nondeterministic-choice construct ``assign such that'' (with syntax borrowed from Dafny \cite{Leino:2010aa}), from \Fig{f1314}, to write
\begin{ProgEqn}[p1102]
	d0,d1\St\ $\PF[d]=(\PF[d0]{+}\PF[d1])/2$ \Com Choose \PF[d0,d1] so that their average is \PF[d].
\end{ProgEqn}%
The analogy with our earlier development is that there the distribution
\PF[d] was specifically $1\,{\PC{p}}\,0$, and we assigned
\[\begin{array}{p{6em}lcl@{~~}l}
	if $\PF[p]\,{\leq}\FF$& \PF[d0],\PF[d1] &=& (\OZ{0}), &(\OZ{2\PF[p]})\\
	if $\PF[p]\,{\geq}\FF$ & \PF[d0],\PF[d1] &=& (\OZ{2\PF[p]{-}1}), &(\OZ{1}) \Qcomma
\end{array}\]
which is a refinement $(\Ref)$ of \Eqn{p1102}.

Our second step is to re-establish the \ProgIL{x\In\ d} -annihilating property that 
\begin{equation}\label{e1433}
	\PF[\Prg{p1102}; d\In\ d0\,{\PCF}\,d1; x\In\ d] \Wide{=} \PF[x\In\ d]\Qcomma
\end{equation}%
which is proved using \WP-calculations agains a general post-expectation \Post, just as before: instead of the assertion \Ass{\PF[p] = (\PF[q]+\PF[r])/2} used at the end of Step 1, we use the assertion \Ass{\PF[d]=(\PF[d0]{+}\PF[d1])/2} established by the assign-such-that.

The third step is again to introduce a loop. But we recall from Step 3 earlier that the loop must be almost-surely terminating and, to show that, we need a variant function. Here we have no \PF[q],\PF[r] that might be set to 0 or 1; we have instead \PF[d0],\PF[d1]. Our variant will be that the ``size'' of one of these distributions must decrease strictly, where we define the \emph{size} of a discrete distribution to be the number of elements to which it assigns non-zero probability.\,%
\footnote{In probability theory this would be the cardinality of its support.}
But our specification \ProgIL{d0,d1\St\ $\PF[d]=(\PF[d0]{+}\PF[d1])/2$} above does not require that decrease; and so we must backtrack in our \CbC\ and make sure that it does.

And we have made an important point: developments following \CbC\ rarely proceed as they are finally presented: the dead-ends are cut off, and only the successful path is left for the audit trail. It highlights the multiple uses of \CbC\ --- that on the one hand, used for teaching, the dead-ends are shown in order to learn how to avoid them; used in production, the successful path remains so that it can be modified in the case that requirements change.\,%
\footnote{And if an error was made in the \CbC\ proofs, the ``successful'' path can be audited to see what the mistake was, why it was made, and how to fix it.}

Thus to establish \AST\ of the loop --that it terminates with probability one-- we strengthen the split of \PF[d] achieved by \ProgIL{d0,d1\St\ $(\PF[d0]{+}\PF[d1])/2 = \PF[d]$} with the decreasing-variant requirement, that either $|\PF[d0]|\,{<}\,|\PF[d]|$ or $|\PF[d1]|\,{<}\,|\PF[d]|$, where we are writing $|{-}|$ for ``size of''. Then the variant $|\PF[d]|$ is guaranteed strictly to decrease with probabililty \FF\ on each iteration. That is we now write
\begin{ProgEqn}[p1442]
	d0,d1\St\quad $(\PF[d0]{+}\PF[d1])/2 = \PF[d] \;\land\; (|\PF[d0]|\,{<}\,|\PF[d]| \lor |\PF[d1]|\,{<}\,|\PF[d]|)$\Qcomma
\end{ProgEqn}%
replacing \Eqn{p1102},
for the nondeterministic choice of \PF[d0] and \PF[d1]. We do not have to re-prove its annihilation property, because the new statement \Eqn{p1442} is a refinement of the \Eqn{p1102} from before (It has a stronger postcondition.) and so preserves all its functional properties. In fact that is the definition of refinement.

\begin{figure}\small
\begin{center}
\hspace*{5em}\includegraphics[width=.8\textwidth]{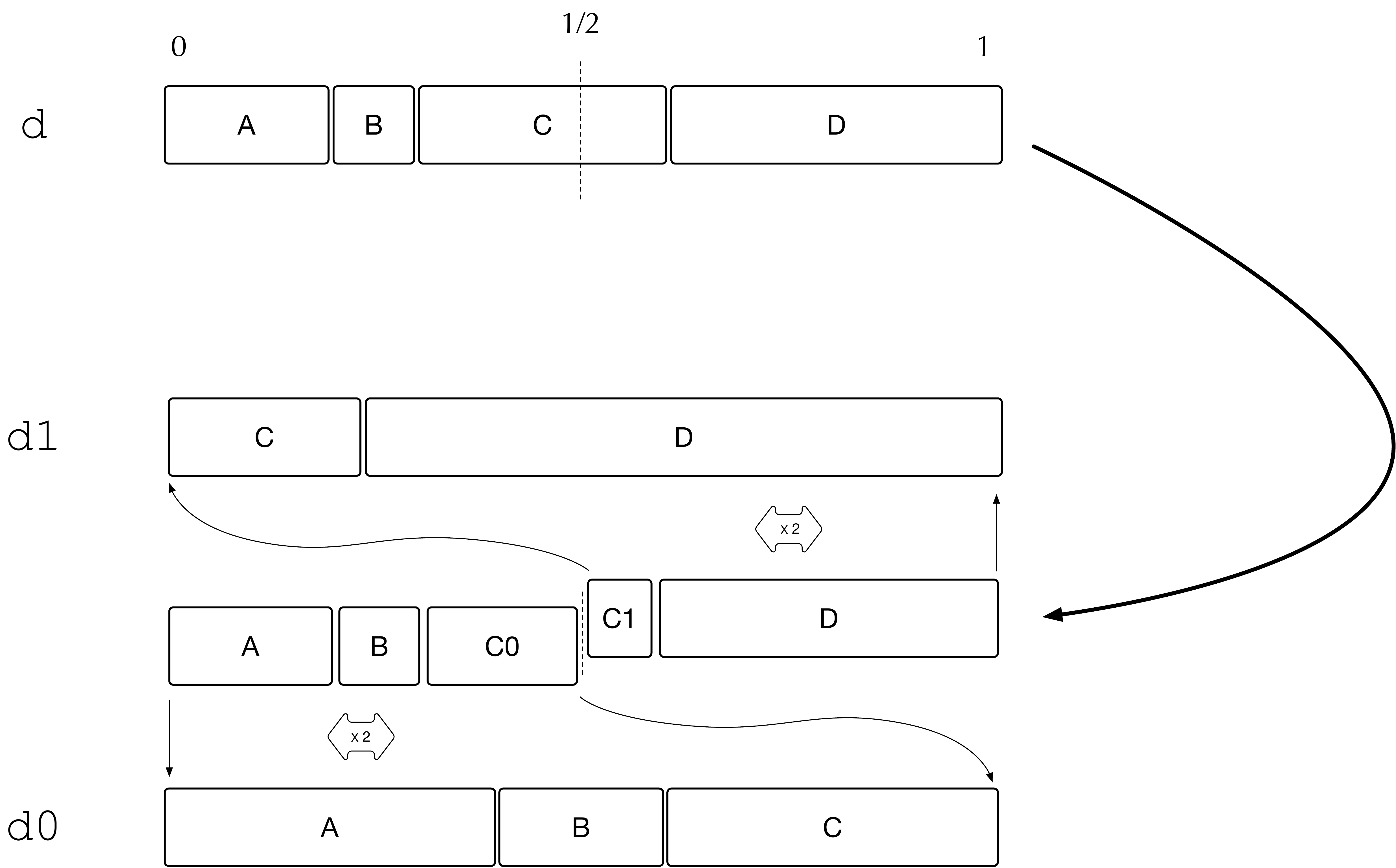}
\end{center}

Suppose that \CalX\ is $\{\textsf{A},\textsf{B},\textsf{C},\textsf{D}\}$, and that the distribution \PF[d] in \CalX\ that we start with is indicated by the size of the rectangles: the size $|\PF{d}|$ of \PF[d] here is therefore 4, because it contains 4 rectangles. We choose $X_0$ to be $\{\textsf{A},\textsf{B},\textsf{C}\}$ and $X_1$ to be $\{\textsf{C},\textsf{D}\}$, so that $X_0{-}X_1$ is  $\{\textsf{A},\textsf{B}\}$ and $X_1{-}X_0$ is  $\{\textsf{D}\}$, and both $\PF[d](X_0{-}X_1)$ and $\PF[d](X_1{-}X_0)$ are no more than \FF. Their overlap is $\{\textsf{C}\}$, whose probability the ``\FF-line'' splits into two pieces: one piece joins \PF[d0] and the other piece joins \PF[d1].

\medskip
Thus by dividing the overall rectangle (representing \CalX\ itself) exactly in the middle, at least one side\,\footnotemark\
must contain strictly fewer than $|\PF{d}|$ rectangles --- and if we double the size of each small rectangle, we get our two distributions \PF[d1] and \PF[d2] such that $\PF[d] = (\PF[d0]+\PF[d1])/2$ and either $|\PF[d0]|\,{<}\,|\PF[d]|$ or $|\PF[d1]|\,{<}\,|\PF[d]|$.

\bigskip
\caption{Dividing a discrete distribution into two pieces}\label{f1511}
\end{figure}

Our next step is to reduce the nondeterminism in \Eqn{p1442} somewhat, choosing a particular way of achieving it: to ``split'' \PF[d] into two parts \PF[d0],\PF[d1] such that the size of at least one part is smaller, we choose two subsets $X_0,X_1$ of \CalX\ whose intersection contains at most one element. That is illustrated in \Fig{f1511}, where $X_0=\{\textsf{A},\textsf{B},\textsf{C}\}$ and $X_1=\{\textsf{C},\textsf{D}\}$. Further, we require that the probabilities $\PF[d](X_0)$ and $\PF[d](X_1)$ assigned by \PF[d] to $X_0{-}X_1$ and $X_1{-}X_0$ are both no more than \FF.\,%
\footnote{Applying \PF[d] to a set means the sum of the \PF[d]-probabilities of the elements of the set.}
Those constraints mean that we can always arrange the subsets so that the ``\FF-line'' of \Fig{f1511} either goes strictly through $X_0\,{\cap}\,X_1$ (if they overlap) or runs between them (if they do not).

\footnotetext{If for example \textsf{C} was much smaller, so that the dividing line went through \textsf{D}, the new distribution \PF[d0] would have support 4, the same as \PF[d] itself. But $|\PF[d1]|$ would have support 1, strictly smaller.}

We then construct \PF[d0] by restricting \PF[d] to just $X_0$, then doubling all the probabilities in that restriction; if they sum to more than 1, we then trim any excess from the one element in $X_0\,{\cap}\,X_1$ that $X_0$ shares with $X_1$. The analogous procedure is applied to generate \PF[d1]. In \Fig{f1511} for example we chose sizes $0.2$, $0.1$, $0.3$ and $0.4$ for the four regions, and the \FF\ line went through the third one. On the left, the $0.2$ and $0.1$ and $0.3$ are doubled to $0.4$ and $0.2$ and $0.6$, summing to $1.2$; thus $0.2$ is trimmed from the $0.6$, leaving $0.4$ assigned to \textsf{C}. The analogous procedure applies on the right.

\subsection{``Decomposition of data into data structures''}
\label{s1451}

The quote is from Wirth \cite{Wirth:71}.
Our program is currently
\begin{ProgEqn}[p1032]
	WHILE |d|$\neq$1 DO \+\\
	d0,d1\St\quad $(\PF[d0]{+}\PF[d1])/2 = \PF[d] \;\land\; (|\PF[d0]|\,{<}\,|\PF[d]| \lor |\PF[d1]|\,{<}\,|\PF[d]|)$ \\	
	d\In\ \PF[d0]\,\PCF\,\PF[d1] \-\\
	OD \\
%	x\St\ d = $\eta$x \Com Recall that $\eta\PF[x]$ is the point distribution on \PF[x].	
	x\In\ d // This is aa trivial choice, because \PF[|d|$=$1] here.
\end{ProgEqn}%
And it is correct: it does refine \PF[x\In\ d] --- but it is rather abstract.
Our next development step will be to make it concrete by realising the distribution-typed variables and the subsets of \CalX\ as ``ordinary'' datatypes using scalars and lists. In correctness-by-construction this is done by deciding, before that translation process begins, what the realisations will be --- and only then is the abstract program transformed, piece by piece. The relation between the abstract- and concrete types is called a \emph{coupling invariant}.
%Partial correctness is preserved, because the nondeterminism has been reduced --- and for example the obviously non-terminating \ProgIL{d0,d1:= d,d} is (now) excluded as an implementation of the choice. The extra conjunct $|\PF[d0]|\,{<}\,|\PF[d]| \lor |\PF[d1]|\,{<}\,|\PF[d]|$ guarantees \AST.

%\section*{(A more general development)}\label{s1511}

Although an obvious approach is to order the type \CalX, say as $x_1,x_2,\ldots,x_N$ and then to realise discrete distributions as lists of length $N$ of probabilities (summing to 1), a more concise representation is suggested by the fact that for example we represent a \emph{two}-point distribution $x_1\PC{p}x_2$ as just \emph{one} number $p$, with the $1{-}p$ implied. Thus we will represent the distribution $p_1,p_2,\ldots\,p_N$ as the list of length $N{-}1$ of ``accumulated'' probabilities: in this case for $p$ we would have a list
\[
	p_1,~~p_1{+}p_2,~~\ldots,~~ \sum_{n=1}^{N-1} p_n\Qcomma
\]
leaving off the $N^{th}$ element of the list since it would always be 1 anyway. Subsets of \CalX\ will be pairs \PF[low],\PF[high] of indices, meaning $\{x_{\PF low},\ldots,x_{\PF high}\}$, and although that can't represent \emph{all} subsets of \CalX, contiguous subsets are all we will need.
Carrying out that transformation gives following concrete version of our abstract program \Prg{p1032} below, where the abstract \PF[d] is represented as the concrete {\PF dL[low:high]}\,, which is the coupling invariant.%
\footnote{The range {\PF [low:high]} is inclusive-exclusive (as in Python). A similar coupling invariant applies to \PF[d0] and \PF[d1]. All three invariants are applied at once.}

\begin{figure}
\begin{ProgEqn}[p1051]
\LCom Discrete distribution \PF[d] in \CalX\ of size $N$ is realised here as \PF[dL] (for ``\PF[d]-list''). \\
	low,high:= 1,$N$ \Com Initial support is all of $\CalX$. \\[0.5ex]
	WHILE low\,$\neq$\,high DO \Com $\PF[low]\,{=}\,\PF[high]$ means support is $\{x_{\PF low}\}$\+\\
	\LCom Current support is $\{x_{\PF[low]},\ldots,x_{\PF[high]}\}$. \\\\
	\LCom Find $X_0$ by examining the probabilities of $x_1, x_2,\ldots$ \\[-.5ex]
	n:= low \Com Determine \PF[dL0] as in \LHS\ of \Fig{f1511}. \\
	WHILE~ n<high$\,\land\,$dL[n]<1/2 ~DO dL0[n]:= 2*dL[n]; n:= n+1 OD \\
	low0,high0:= low,n \Com Subset $X_0$ is $\{x_{\PF[low0]},\ldots,x_{\PF[high0]}\}$\,. \\\\
	\LCom Find $X_1$ by examining the probabilities of $x_N, x_{N-1},\ldots$ \\[-.5ex]
	n:= high-1 \Com Determine \PF[dL1] as in \RHS\ of \Fig{f1511}. \\
	WHILE~ low$\leq$n$\,\land\,$1/2<dL[n] ~DO dL1[n]:= 2*dL[n]-1; n:= n-1 OD \\
	low1,high1:= n+1,high \Com Subset $X_1$ is $\{x_{\PF[low1]},\ldots,x_{\PF[high1]}\}$\,. \\\\
	\LCom Use fair coin to choose between \PF[dL0] and \PF[dL1].\\
	(dL,low,high)\In\ (dL0,low0,high0)\,\PCF\;(dL1,low1,high1) \-\\\\
	OD \\
	x:= $x_{\PF[low]}$ \Com Extract sole element of point distribution's support.
\end{ProgEqn}%
\caption{Implement any discrete choice using only a fair coin.}\label{f0834}
\end{figure}
And in \Prg{p1051} of \Fig{f0834} we have, finally, a concrete program that can actually be run. Notice that it has exactly the same \emph{structure} as \Prg{p1032}: split (the realisations of) \PF[d] into \PF[d0] and \PF[d1]; overwrite \PF[d] with one of them; exit the loop when \PF[|d|] is one.

Neverthess, as earlier in \Sec{step 5}, further development steps might still be possible now that everything is together in one place:\,%
\footnote{Note the necessity of keeping this as two steps: first data-refine, then (if you can) optimise algorithmically.}
and indeed, recognising that only one of \PF[dL0],\PF[dL1] will be \emph{used}, we can rearrange \Prg{p1051}'s body so that only that only one of them will be \emph{calculated} --- and it can be updated as we go. That gives 
our really-final-this-time program \Eqn{p1511} in \Fig{f1541}, which will -without further intervention-- use a fair coin to choose a value $x_n$ according to \emph{any} given discrete distribution $d$ on finite \CalX. Its expected number of coin flips is no worse than $2N{-}2$, where $N$ is the size of \CalX, thus agreeing with expected 2 flips for the program \Eqn{p0955} in \Sec{step 5} that dealt with the simpler case $d = (1\PC{p}0)$ where \CalX\ was $\{1,0\}$.
%\Footnote{Justify this; compare it with the simple case of a binary distribution.}

\begin{figure}
\begin{ProgEqn}[p1511]
\LCom Assume discrete distribution $d$ over $\CalX=\{x_1,\ldots,x_N\}$ of size $N$ \\[-0.3em]
\LCom has been represented cumulatively in list \PF[dL], as described above. \\[2ex]
	low,high:= 1,$N$ \Com Initial support is all of $\CalX$. \\
	WHILE low\,$\neq$\,high DO \Com $\PF[low]\,{=}\,\PF[high]$ means support is $\{x_{\PF low}\}$\+\\[0.5ex]
		\LCom Fair coin flipped here. (Recall \Fig{f1202}.) \\
		IF 1/2 THEN \Com Then update \PF[dL] as in \LHS\ of \Fig{f1511}. \+\\
			n:= low \\
			WHILE~ n<high$\,\land\,$dL[n]<1/2~DO dL[n]:= 2*dL[n]; n:= n+1 OD \\
			high:= n \-\\
		ELSE \Com Else update \PF[dL] as in \RHS\ of \Fig{f1511}. \+\\
			n:= high-1 \\
			WHILE~ low$\leq$n$\,\land\,$1/2<dL[n]~\,DO dL[n]:= 2*dL[n]-1; n:= n-1 OD \\
			low:= n+1 \-\\
		FI \-\\[0.5ex]
	OD \\
	x:= $x_{\PF[low]}$ \Com Extract sole element of point distribution \PF[dL]'s support.
\end{ProgEqn}
\caption{Optimisation of \Prg{p1051}\label{f1541}}
\end{figure}

It's again worth emphasising --because it is the main point-- that the correctness arguments for all of these steps are isolated from each other: in \CbC\ every step's correctness is determined by looking at that step alone.  Thus for example nothing in the translation process just above involved reasoning about the earlier steps, whether \Prg{p1032} actually implemented the \ProgIL{x\In\ d} that we started with: we didn't care, and we didn't check. We just translated \Prg{p1032} into \Prg{p1051} regardless. And the subsequent rearrangement of  \Eqn{p1051} into \Prg{p1511} similarly made no use of \Prg{p1051}'s provenence. 

All that is to be contrasted with the more common approach in which \emph{only} intuition (and experience, and skill) is used, in which our final \Prg{p1511} might be written all at once at this concrete level, only then checking (testing, debugging, hoping) afterwards that our intuitions were correct. A transliteration of \Prg{p1511} into Python is given in App.\,\ref{a1549}.
\begin{figure}\small
\vspace{0ex}
\begin{center}
\includegraphics[width=\textwidth]{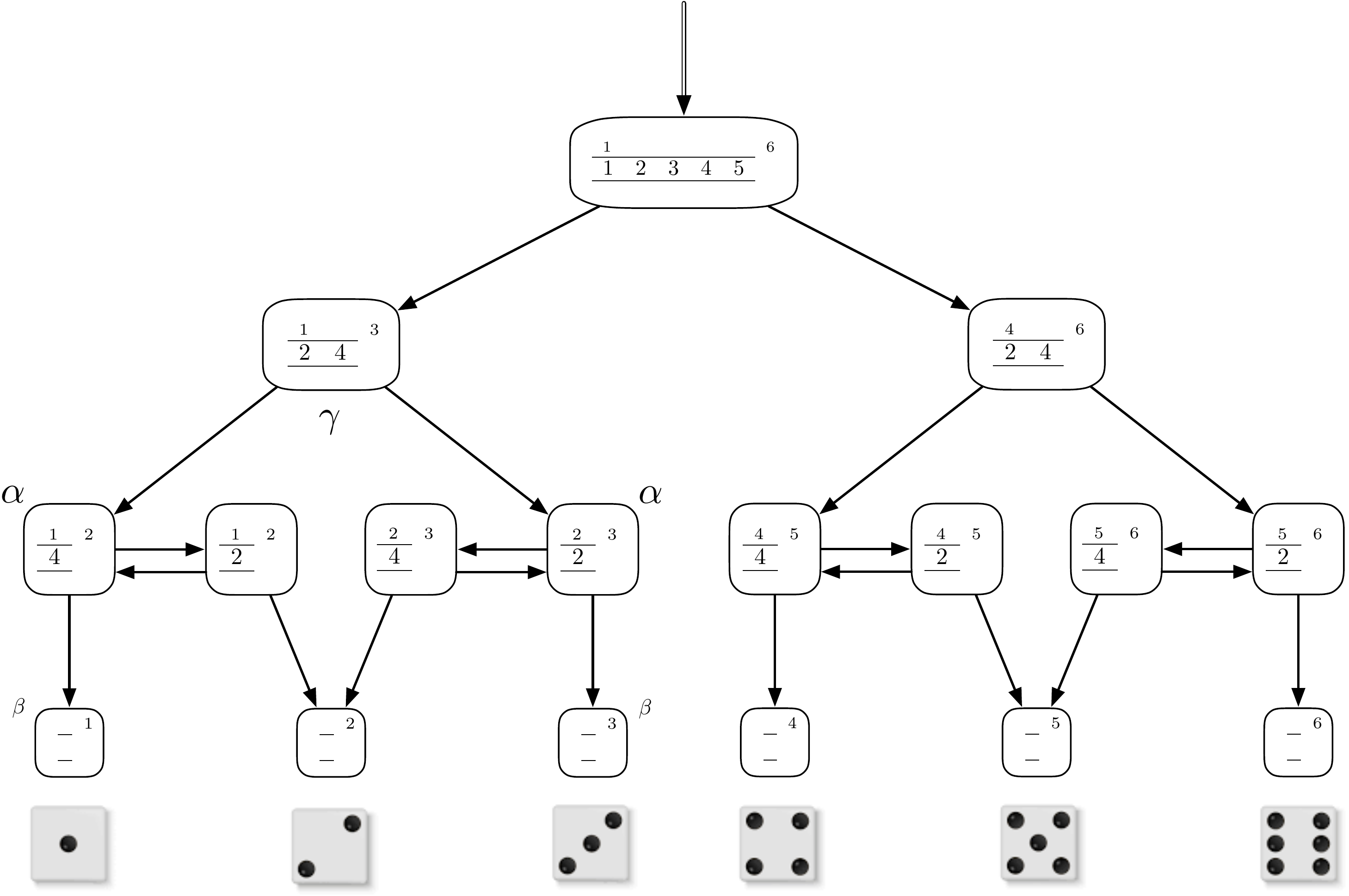}
\end{center}

\bigskip
Each interior node has two possible successors chosen with equal probability, and each final-die node is reached with the same probability $\NF{1}{6}$. There are 17 nodes, and the expected number of coin flips is 4. 

The nodes' origins are shown by labelling them with \PF[low], \PF[d] and \PF[high] from the states in the generating program that gave rise to them, representing the current probability distribution \PF[d] yet to be realised over over the remaining subset $\{\PF[low],\ldots,\PF[high]\}$ of possible results. With probabilities normalised out of 6 for neatness, a typical label is
\[
	\begin{array}{llllll}
	^{\PF[low]}&&&&&^{\PF[high]}
	\\[-.5em]\cline{1-5}
	\multicolumn{5}{c}{\longleftarrow6{\times}\PF[d]\longrightarrow}&\Qcomma
	\\[-.1em]\cline{1-5}
	\end{array}
\]

\smallskip\noindent where we recall that $\PF[d]$ gives the \emph{sum} of the probabilities for $x_{\PF{low}},x_{\PF{low}+1},\ldots,x_{\PF{high}-1}$ and that \PF[d] for $x_{\PF[high]}$ is left out, because it is always 1. Thus for example $\PF[low]=2$ and $\PF[high]=3$ and $\PF[d]=[4]$ represents the distribution over support $\{2,3\}$ of $\NF{4}{6}$ for 2 and $1{-}\NF{4}{6}$ for 3, that is $2\PCNF{2}{3}3$.

\begin{center}\raisebox{.5ex}{\rule{20em}{.5pt}}\end{center}\vfill

The well-known (optimal) algorithm of Knuth and Yao for simulating a die with a fair coin has 13 states and $\NF{11}{3}$ expected coin flips \cite{Knuth:1976aa} --- and it can be obtained from here by one last correctness-preserving step. Eliminate the choice $\gamma$, so that the two $\alpha$ and the two $\beta$ nodes are merged; since that also merges the two die-rolls 1 and 3, restore the $\gamma$ choice as a new fair choice $\gamma'$ over $\{1,3\}$, just below the merged $\beta$'s. (The nodes leading to die-roll 2 are merged as well, but it makes no difference.)

Concentrating on the left (justified by symmetry), we see that the original $\gamma$ choice must be done every time; but its replacement $\gamma'$ is done only $\NF{2}{3}$ of the time. That realises exactly the $\NF{1}{3}$ efficiency advantage that Knuth/Yao optimal algorithm has over the one synthesised here by our general \Prg{p1511}.

\medskip\caption{Simulating a fair die with a fair coin}\label{f1514}
\end{figure}

\section{An everyday application: \\ simulating a fair die using only a fair coin}
\label{s1512}

\Prg{p1511} of the previous section works for any discrete distribution, without having to adapt the program in any way. However if the distribution's probabilities are not too bizarre, then the number of different values for \PF[low] and \PF[d] and \PF[high] might be quite small --- and then the program's behaviour for that distribution in particular can be set out as a small probabilistic state machine.

In \Fig{f1511} we take \PF[d] to be the uniform distribution over the possible die-roll outcomes $\{1,2,3,4,5,6\}$, and show the state machine that results.
For that state machine in particular, we propose one last correctness-preserving step: it takes us to the optimal die-roll algorithm of Knuth and Yao \cite{Knuth:1976aa}.

\section{Why was this ``correctness by construction''?}
The programs here are not themselves remarkable in any way. (The optimality of the Knuth/Yao algorithm is not our contribution). Even the mathematical insights used in their construction are well known, examples of elementary probability theory. \CbC\ means however applying those insights in a systematic, layered way so that the reasoning in each layer does not depend on earlier layers, and does not affect later ones. The steps were specifically
{\begin{enumerate}
\item Start with the \emph{specification} \PF[x\In\ d] at the beginning of \Sec{s1041}.
\item Prove a one-step annihilation property \Eqn{e1433} for that specification.
\item Use a general loop rule to prove loop-annihilation \Prg{p1032}, after Strengthening \Prg{p1102} to \Prg{p1442} to establish \AST.
\item Propose strategy \Fig{f1511} for the loop body of \Prg{p1032}.
\item Propose data representation of finite discrete distributions as lists, in \Sec{s1451}, realising the strategy of \Fig{f1511} in the code of \Prg{p1051}.\
\item Rearrange \Prg{p1051} to produce a more efficient final program \Prg{p1511}.
\item[]
\item Note that \textbf{correctness-by-construction guarantees} that \Prg{p1511} refines \PF[x\In\ d]
	for any \PF[d].
\item[]
\item Apply \Prg{p1511} to the fair die, to produce state chart of \Fig{f1514}.
\item Modify \Fig{f1514} to produce the Knuth/Yao (optimal) algorithm \cite{Knuth:1976aa}.
\item[]
\item Note that \textbf{correctness-by-construction guarantees} that the Knuth/Yao (optimal) algorithm implements a fair die.
\end{enumerate}}

\CbC\ also means that since all those steps are done explicitly and separately, they can be checked easily as you go along, and audited afterwards. But to apply \CbC\ effectively, and \emph{honestly}, one must have a rigorous semantics that justifies every single development step made. In our example here, that was supplied here by the semantics of \pGCL\ \cite{McIver:05a}. But working in any ``wide spectrum'' language, right from the (abstract) start all the way to the (concrete) finish, means that many of those rigorous steps can be checked by theorem provers.

\newpage
%\bibliographystyle{plain}
%\bibliography{CCMbib200709}

\begin{thebibliography}{10}

\bibitem{Dijkstra:aa}
Edsger~W Dijkstra.
\newblock On the reliability of programs ({EWD}303).

\bibitem{Dijkstra:76}
Edsger~W. Dijkstra.
\newblock {\em A Discipline of Programming}.
\newblock Prentice-Hall, 1976.

\bibitem{Floyd:67}
R.W. Floyd.
\newblock Assigning meanings to programs.
\newblock In J.T. Schwartz, editor, {\em Mathematical Aspects of Computer
  Science}, number~19 in Proc Symp Appl Math., pages 19--32. American
  Mathematical Society, 1967.

\bibitem{Hoare:69}
C.~A.~R. Hoare.
\newblock An axiomatic basis for computer programming.
\newblock {\em Commun. {ACM}}, 12(10):576--580, 1969.

\bibitem{Jones:89}
C.~Jones and G.~Plotkin.
\newblock A probabilistic powerdomain of evaluations.
\newblock In {\em Proceedings of the {IEEE} 4th Annual Symposium on Logic in
  Computer Science}, pages 186--95, Los Alamitos, Calif., 1989. Computer
  Society Press.

\bibitem{Knuth:1976aa}
D.~Knuth and A.~Yao.
\newblock {\em Algorithms and Complexity: New Directions and Recent Results},
  chapter The complexity of nonuniform random number generation.
\newblock Academic Press, 1976.

\bibitem{Kozen:81}
D.~Kozen.
\newblock Semantics of probabilistic programs.
\newblock {\em Jnl Comp Sys Sci}, 22:328--50, 1981.

\bibitem{Kozen:83}
D.~Kozen.
\newblock A probabilistic {PDL}.
\newblock In {\em Proceedings of the 15th ACM Symposium on Theory of
  Computing}, pages 291--7, New York, 1983. ACM.

\bibitem{Leino:2010aa}
K.R.M. Leino.
\newblock Dafny: An automatic program verifier for functional correctness.
\newblock In Voronkov~A. Clarke~E.M., editor, {\em Logic for Programming,
  Artificial Intelligence, and Reasoning. LPAR 2010.}, volume 6355 of {\em
  Lecture Notes in Computer Science}. Springer, 2010.

\bibitem{McIver:05a}
Annabelle McIver and Carroll Morgan.
\newblock {\em Abstraction, Refinement and Proof for Probabilistic Systems}.
\newblock Monographs in Computer Science. Springer, 2005.

\bibitem{McIver:2017aa}
Annabelle McIver, Carroll Morgan, Benjamin~Lucien Kaminski, and Joost-Pieter
  Katoen.
\newblock A new proof rule for almost-sure termination.
\newblock {\em Proc. ACM Program. Lang.}, 2(POPL), December 2017.

\bibitem{Morgan:96d}
Carroll Morgan, Annabelle McIver, and Karen Seidel.
\newblock Probabilistic predicate transformers.
\newblock {\em ACM Trans. Program. Lang. Syst.}, 18(3):325--353, May 1996.

\bibitem{Morgan:96b}
C.C. Morgan.
\newblock Proof rules for probabilistic loops.
\newblock In He~Jifeng, John Cooke, and Peter Wallis, editors, {\em Proc
  BCS-FACS 7th Refinement Workshop}, Workshops in Computing. Springer, July
  1996.
\newblock \texttt{http://www.bcs.org/upload/pdf/ewic\underline{
  }rw96\underline{ }paper10.pdf}.

\bibitem{Vazsonyi:2002aa}
Andrew Vazsonyi.
\newblock {\em Which Door has the Cadillac? Adventures of a Real-Life
  Mathematician}.
\newblock Writers Club Press, 2002.

\bibitem{Wirth:71}
N.~Wirth.
\newblock Program development by stepwise refinement.
\newblock {\em Comm ACM}, 14(4):221--7, 1971.

\end{thebibliography}

\newpage\appendix\section{\Prg{p1511} implemented in Python}
\label{a1549}
{\small %1207
\begin{verbatim}
#   Run 1,000,000 trials on a fair-die simulation.
#
#   bash-3.2$ python ISoLA.py
#   1000000
#   1 1 1 1 1 1
#   Relative frequencies
#         0.998154 1.00092  0.996474 0.998664 1.004928  1.00086
#   realised, using 4.001938 flips on average.
\end{verbatim}
\begin{verbatim}
import sys
from random import randrange

# Number of runs, an integer on the first line by itself.
runs = int(sys.stdin.readline())

# Discrete distribution unnormalised, as many subsequent integers as needed.
# Then EOT.
d= []
for line in sys.stdin.readlines():
    for word in line.split(): d.append(int(word))
sizeX= len(d) # Size of initial distribution's support.

# Construct distribution's representation as accumulated list dL_Init.
# Note that length of dL_Init is sizeX-1,
#    because final (normalised) entry of 1 is implied.
# Do not normalise, however: makes the arithmetic clearer.
sum,dL_Init= d[0],[]
for n in range(sizeX-1): dL_Init= dL_Init+[sum]; sum= sum+d[n+1]

tallies= []
for n in range(sizeX): tallies= tallies+[0]

allFlips= 0 # For counting average number of flips.
for r in range(runs): flips= 0
    
### Program (14) proper starts on the next page.
\end{verbatim}
\newpage
\begin{verbatim}    
    ### Program (14) starts here.
    low,high,dL= 0,sizeX-1,dL_Init[:] # Must clone dL_Init.
    # print "Start:", low, dL[low:high], high

    while low<high:
        flip= randrange(2) # One fair-coin flip.
        flips= flips+1

        if flip==0:
            n= low
            while n<high and 2*dL[n]<sum: dL[n]= 2*dL[n]; n= n+1
            high= n # Implied dL0[high]=1 performs trimming automatically.
            # print "Took dL0:", low, dL[low:high], high # dL0 has overwritten dL.

        else: # flip==1
            n= high-1
            while low<=n and 2*dL[n]>sum: dL[n]= 2*dL[n]-sum; n= n-1
            low= n+1 # Implied dL1[low]=0 performs trimming automatically.
            # print "Took dL1", low, dL[low:high], high # dL1 has overwritten dL.

    # print "Rolled", low, "in", flips, "flips."
    ### Program (14) ends here.
    
    tallies[low]= tallies[low]+1
    allFlips= allFlips+flips
    
print "Relative frequencies"
for n in range(sizeX): print "     ", float(tallies[n])/runs * sum
print "realised, using", float(allFlips)/runs, "flips on average."
\end{verbatim}
} % 1207

\end{document}